\documentclass[prd,nofootinbib,english]{revtex4}
\usepackage{csvsimple}
\usepackage{amsmath}
\usepackage{amsfonts,color}
\usepackage{amssymb,float,multirow}
\usepackage[utf8]{inputenc}
\usepackage{longtable}
\usepackage{appendix}
\usepackage{color}
\usepackage[unicode=true,pdfusetitle,
 bookmarks=true,bookmarksnumbered=true,bookmarksopen=true,bookmarksopenlevel=3,
 breaklinks=false,pdfborder={0 0 1},backref=false,colorlinks=true]
 {hyperref}
\usepackage{enumitem}

\usepackage{tikz}
\usetikzlibrary{shapes.geometric}
\usetikzlibrary{arrows.meta,arrows}

\begin{document}

\title{Constraints on Cubic and $f(P)$ Gravity from the Cosmic Chronometers, BAO \& CMB datasets : Use of Machine Learning Algorithms }
\author{Kinsuk Giri}
\email{kinsuk@nitttrkol.ac.in}
\affiliation{Department of CSE, National Institute of Technical Teacher's Training and Research, Block-FC, Sector-III, Salt Lake, Kolkata-700 106, India}

\author{Prabir Rudra}
\email{prudra.math@gmail.com, prabir.rudra@asutoshcollege.in}
\affiliation{Department of Mathematics, Asutosh College, Kolkata-700 026, India}


\begin{abstract}
In this work we perform an observational data analysis
on Einsteinian cubic gravity and $f(P)$ gravity with the objective of constraining the parameter space of the theories. We use the 30 point $z-H(z)$ cosmic chronometer data as the observational tool for our analysis along with the BAO and the CMB peak parameters. The $\chi^2$ statistic is used for the fitting analysis and it is minimized to obtain the best fit values for the free model parameters. We have used the Markov chain Monte Carlo algorithm to obtain bounds for the free parameters. To achieve this we used the  publicly available \textit{CosmoMC} code to put parameter bounds and subsequently generate contour plots for them with different confidence intervals. Besides finding the Hubble parameter $H$ in terms of the redshift $z$ theoretically from our gravity models, we have exercised correlation coefficients and two \textit{machine learning} models, namely the linear regression (LR) and artificial neural network (ANN), for the estimation of $H(z)$. For this purpose, we have developed a \textit{Python} package for finding the parameter space, performing the subsequent statistical analysis and prediction analysis using machine learning. We compared both of our theoretical and estimated values of $H(z)$ with the observations. It is seen that our theoretical and estimated models from machine learning  performed significantly well when compared with the observations.
\end{abstract}

\maketitle

\section{Introduction}
Since the discovery of the late cosmic acceleration (Riess et al. 1998; Perlmutter et al. 1999; Spergel et al. 2003), General Relativity (GR) suffered a big set-back due to its incompatibility and inability to explain the event. Since then modified gravity theories (Nojiri \& Odintsov 2007; Nojiri, Odintsov \& Oikonomou 2017) have become a useful tool in describing the acceleration of the universe. The basic aim of a modified gravity theory is to comprehensively describe the history of the universe staring from the early inflationary phase to the late time acceleration, always complying with the observations. Modification of Einstein gravity basically involves the modification of the Einstein-Hilbert (EH) action, by generalizing the gravity Lagrangian $R$ (where $R$ is the Ricci scalar invariant given by $R=g^{\mu\nu}R_{\mu\nu}$). Corrections to EH action may be inflicted from higher order curvature terms, invariants coming from the matter sector, inclusion of torsion in the theory, etc. This can motivate one to search for higher-order corrections to the EH term, thus resulting in higher-order gravity theories. It should be stated here, that the higher-order corrections are motivated from the fact that such terms naturally arise in the effective action of a complete string theory (Gross \& Sloan 1987) resulting in a renormalizable and hence a quantizable theory of gravitation (Stelle 1977). Some examples include topologically massive gravity (Deser, Jackiw \& Templeton 1982, 2000), new massive gravity in three dimensions (Bergshoeff, Hohm \& Townsend 2009) and critical gravity (Lu \& Pope 2011). The whole concept of holography has in fact motivated the construction of higher-order theories like quasitopological gravity (Oliva \& Ray 2010, Myers, Paulos \& Sinha 2010). One important aspect of higher order theories is that some of them are equivalent to Einstein gravity at the linearized level in vacuum. In these cases the only physical mode transmitted by the metric perturbation is a transverse and massless graviton. Certain examples include the quasitopological gravity (Oliva \& Ray 2010, Myers, Paulos \& Sinha 2010) and certain f(Lovelock) theories (Lovelock 1971, Bueno et al. 2016, Karasu, Kenar \& Tekin 2016).

It is known that the most fundamental generalization of the EH action is inflicted by replacing $R$ by an arbitrary function $f(R)$ resulting in $f(R)$ theories (De Felice \& Tsujikawa 2010; Sotiriou \& Faraoni 2010). Inclusion of an arbitrary function of $R$ in the field equations help us to probe the non-linear effects of the curvature invariant. Significant development in $f(R)$ theory can be found in Nojiri \& Odintsov (2006), Sotiriou (2006), Amendola, Polarski \& Tsujikawa (2007), Song, Hu \& Sawicki (2007), Rudra (2016), Rudra (2020). Including the second order corrections of the curvature tensor in EH action one can formulate the $f(G)$ theory where $G=R^{2}-R_{\mu\nu}R^{\mu\nu}+R_{\mu\nu\alpha\beta}R^{\mu\nu\alpha\beta}$ is the Gauss-Bonnett invariant (Nojiri \& Odintsov 2005; Nojiri, Odintsov \& Sasaki 2005; De. Felice \& Tsujikawa 2009). Important theoretical advancement in $f(G)$ gravity may be found in Goheer et al. (2009), Rudra (2014), Shamir \& Naz (2020). Other examples of modified gravity theories are Weyl gravity (Flanagan 2006) arising from non-metricity, Lovelock gravity (Lovelock 1971; Deruelle \& Farina-Busto 1990), scalar tensor theories like Brans-Dicke (Brans \& Dicke 1961) \& Galileon gravity (Nicolis, Rattazzi \& Trincherini 2009; Deffayet, Esposito-Farese \& Vikman 2009, Leon \& Saridakis 2013, Rudra, Faizal \& Ali 2016), theories based on torsion such as $f(T)$ (Bengochea \& Ferraro 2009; Linder 2010; Chen et al. 2011; Jamil et al. 2012; Bahamonde, Marciu \& Rudra 2018) and $f(T,T_{G})$ theories (Kofinas \& Saridakis 2014a; Kofinas \& Saridakis 2014b; Bahamonde et al. 2021), etc. Further developments in these theories may be found in Rudra \& Debnath (2014), Koyama (2020), Song, Zhang \& Ma (2020), Rudra, Biswas \& Debnath (2014), Rudra \& Maity (2018).

In spite of being highly motivated, modified gravity theories have several disadvantages. Since the coupling of the curvature invariants depend of the dimensions of the spacetime, these eventually boil down to different theories in different dimensions. The higher order curvature terms in the action may give rise to field equations whose order is greater than the second order. This may not always indicate instabilities or pathologies, but yet it is always a matter of concern.

A very interesting higher order theory known as the Einsteinian cubic gravity (ECG) (Bueno \& Cano 2016a) was developed recently where the authors used cubic contraction of the Riemann tensor $R_{\mu\nu\alpha\beta}$ to generate the cubic invariant. In the formulation the authors used the linearization technique of general higher order gravities. Although being highly non-linear the theory contains non-topological terms responsible for basic health conditions. It was shown that the theory admits spherically symmetric black hole solutions with a second order differential equation for the metric function (Bueno \& Cano 2016b; Hennigar \& Mann 2017). The cosmology of ECG was studied and it was found that it has a mechanism that triggers an early inflation and a late time acceleration close to the $\Lambda$CDM behaviour (Arciniega, Edelstein \& Jaime 2020). Observing these developments, the theory was further extended to $f(P)$ gravity ($P$ being the invariant from ECG) where the gravity Lagrangian was given as the Ricci scalar coupled with an arbitrary function of the cubic invariant $P$ (Erices, Papantonopoulos \& Saridakis 2019).
The theory was formulated in four dimensions and the cosmology of the model was studied. It was seen that $f(P)$ gravity admitted early inflation and late acceleration even with a vanishing cosmological constant. A complete dynamical system analysis on $f(P)$ gravity was performed by Marciu (2020). Recently the $f(P)$ gravity is further extended to $f(R,P)$ gravity in Marciu (2021), where the author studied the dark effects of the theory.

Inspired by the success of ECG and $f(P)$ theories in the cosmological context, we want to conduct an observational data analysis on both the theories to constrain their parameter space. The most important aspect of a theoretical model is its degree of compliance with the observations. Fitting the theory with the observational data one can obtain bounds for the free parameters of the model using various statistical tools. This is a vital exercise for any theory and with the bounds on the model parameters the theory gains physical viability. In this work we will use the cosmic chronometer data from the slowly evolving distant galaxies as our observational tool. We will use various statistical procedures like the $\chi^2$ minimization technique, Markov chain Monte Carlo method and machine learning algorithms for our analysis. The paper is organized as follows: In section II we discuss the basic equations of ECG and $f(P)$ gravity. In section III we talk about observational data analysis with the cosmic chronometer data coupled to Baryon acoustic oscillation and Cosmic microwave background peak parameters using the $\chi^{2}$ minimization technique and the Markov chain Monte Carlo algorithm. In section IV we discuss the statistical analysis and machine learning methodology. In section V we present all the results obtained from section III \& IV. Finally the paper ends with a conclusion in section VI.

\section{Basic equations of cubic and $f(P)$ gravity}
The basic way of imposing modifications to Einstein's theory of
General Relativity (GR) is by introducing new scalar invariants in
the gravitational Lagrangian of the Einstein-Hilbert action that
will replace the orthodox Ricci scalar $R$. Now there can be many
such possibilities from the mathematical point of view. We can
always play around with the Riemann tensor $R_{\mu\nu\rho\sigma}$,
Ricci tensor $R_{\mu\nu}$, energy momentum tensor $T_{\mu\nu}$,
etc. and explore their possible contractions and produce scalars
from such exercises. Although their physical significance and
importance to cosmology is a totally different question and needs
thorough study, their mathematical significance is unquestionable.
Nevertheless while undertaking such an exercise one would like to
be in compliance with three rules: (i) The theory possess an
identical spectrum as GR.  (ii) It should be non-topological in
nature and should possess non-trivial terms in four dimensions.
(iii) The field equations retrieved from such a theory must be of
second order.

In 4 dimensional spacetime a general non-topological cubic term
would be given by (Erices, Papantonopoulos \& Saridakis 2019),

$$P=\beta_{1}R_{\mu~~\nu}^{~~\rho~~\sigma}R_{\rho~~\sigma}^{~~\gamma~~\delta}R_{\gamma~~\delta}^{~~\mu~~\nu}
+\beta_{2}R_{\mu\nu}^{\rho\sigma}~R_{\rho\sigma}^{\gamma\delta}~R_{\gamma\delta}^{\mu\nu}+\beta_{3}R^{\sigma\gamma}R_{\mu\nu\rho\sigma}{R^{\mu\nu\rho}}_{\gamma}
+\beta_{4}RR_{\mu\nu\rho\sigma}R^{\mu\nu\rho\sigma}$$
\begin{equation}\label{cubicterm}
+\beta_{5}R_{\mu\nu\rho\sigma}R^{\mu\rho}R^{\nu\sigma}+\beta_{6}R^{\nu}_{\mu}R^{\rho}_{\nu}R^{\mu}_{\rho}+\beta_{7}R_{\mu\nu}R^{\mu\nu}R+\beta_{8}R^{3}
\end{equation}
where $\beta_{i}$ are parameters. If the condition (i) stated
above is to be satisfied then we should have,
\begin{equation}\label{cond1}
\beta_{7}=\frac{1}{12}\left(3\beta_{1}-24\beta_{2}-16\beta_{3}-48\beta_{4}-5\beta_{5}-9\beta_{6}\right)
\end{equation}
\begin{equation}\label{cond2}
\beta_{8}=\frac{1}{72}\left(-6\beta_{1}+36\beta_{2}+22\beta_{3}+64\beta_{4}+3\beta_{5}+9\beta_{6}\right)
\end{equation}

The action for the cubic gravity is given by,
\begin{equation}\label{actioncubic}
\mathcal{S}=\int \sqrt{-g}d^{4}x\left[\frac{R}{2\kappa}+\alpha P\right]+\int \left(\mathcal{L}_{m}+\mathcal{L}_{rad}\right)\sqrt{-g}d^{4}x
\end{equation}
where $P$ is the invariant defined in eqn.(\ref{cubicterm}) and
$\alpha$ is the coupling parameter. $\kappa=8\pi G$ is the Newton's constant and we have considered a vanishing cosmological constant. $\mathcal{L}_{m}$ is the matter Lagrangian and $\mathcal{L}_{rad}$ is the Lagrangian for the pressure-less radiation. Varying the above action (\ref{actioncubic}) with respect to the metric we arrive at the following field equations,
\begin{equation}\label{efe1}
G_{\mu\nu}=\kappa \left(T_{\mu\nu}+\alpha H_{\mu\nu}+T_{\mu\nu}^{rad}\right)
\end{equation}
where the energy-momentum tensors $T_{\mu\nu}$ and $T_{\mu\nu}^{rad}$ are given by,
\begin{equation}\label{emt1}
T_{\mu\nu}=-\frac{2}{\sqrt{-g}}\frac{\delta \left(\sqrt{-g}\mathcal{L}_{m}\right)}{\delta g^{\mu\nu}}~,~~~~~~~~~~~~T_{\mu\nu}^{rad}=-\frac{2}{\sqrt{-g}}\frac{\delta \left(\sqrt{-g}\mathcal{L}_{rad}\right)}{\delta g^{\mu\nu}}
\end{equation}
and
\begin{equation}
H_{\mu\nu}=-\frac{2}{\sqrt{-g}}\frac{\delta \left(\sqrt{-g}P\right)}{\delta g^{\mu\nu}}=g_{\mu\nu}P+R^{\alpha \beta \rho}{}_{(\mu}K_{\nu)\rho \alpha \beta}+2\nabla^{\alpha}\nabla^{\beta}K_{\alpha(\mu\nu)\beta}
\end{equation}
where $\nabla_{\mu}$ is the covariant derivative with respect to the metric $g_{\mu\nu}$. The tensor $K_{\alpha\beta\mu\nu}$ is defined by,

$$K_{\alpha\beta\mu\nu}=\frac{\partial P}{\partial R^{\alpha\beta\mu\nu}}$$
\begin{equation}\label{Kten}
=12\left(\frac{1}{2}R_{\alpha\beta}^{~~~\rho\sigma}R_{\mu\nu\rho\sigma}+6R_{\alpha}^{~\rho}{}_{[\mu}{}^{\sigma} R_{\nu]\sigma\beta\rho}+2g{}_{\beta[\mu}R_{\nu]\sigma\alpha\rho}R^{\rho\sigma}-2g{}_{\alpha[\mu}R_{\nu]\sigma\beta\rho}R^{\rho\sigma}-4R_{\rho}{}_{[\mu}g_{\nu]}{}_{[\alpha}R_{\beta]}~^{\rho}-2R_{\alpha[\mu}R_{\nu]\beta}\right)
\end{equation}
Here the tensor $H_{\mu\nu}$ arises from the contribution of the invariant $P$.

In order to explore the cosmological implications of the cubic gravity we consider a homogeneous and isotropic spacetime given by the following Friedmann-Lemaitre-Robertson-Walker (FLRW) metric,
\begin{equation}\label{flrw1}
ds^{2}=-dt^{2}+a^{2}(t)\delta_{ij}dx^{i}dx^{j}
\end{equation}
where $a(t)$ is the scale factor and $\delta_{ij}$ is the Kronecker delta. The matter sector will be represented by perfect a fluid with the energy momentum tensor
\begin{equation}\label{emt2}
T_{\mu\nu}=\left(\rho_{m}+p_{m}\right)u_{\mu}u_{\nu}+p_{m}g_{\mu\nu}
\end{equation}
where $\rho_{m}$ and $p_{m}$ are respectively the energy density and pressure of matter and $u_{\mu}$ is the four-velocity of the fluid. Equipped with all these considerations we can now write the FLRW equations from the field equations (\ref{efe1}) as,
\begin{equation}\label{frw1}
3H^{2}=\kappa\left(\rho_{m}+6\alpha \tilde{\beta}H^{6}+\rho_{rad}\right)
\end{equation}

\begin{equation}\label{frw2}
3H^{2}+2\dot{H}=-\kappa\left[p_{m}-6\alpha \tilde{\beta}H^{4}\left(H^{2}+2\dot{H}\right)\right]
\end{equation}
where $H=\frac{\dot{a}}{a}$ is the Hubble parameter and dots (.) represent derivatives with respect to time. Here we have defined the parameter $\tilde{\beta}$ as,
\begin{equation}\label{tbeta}
\tilde{\beta}\equiv -\beta_{1}+4\beta_{2}+2\beta_{3}+8\beta_{4}   
\end{equation}
Moreover the condition for the second order field equations is
satisfied if we have,
\begin{equation}\label{cond3}
\beta_{6}=4\beta_{2}+2\beta_{3}+8\beta_{4}+\beta_{5}
\end{equation}
It should be noted that under the FLRW geometry considering the equations (\ref{cond1}) and (\ref{cond3}), the cubic invariant can be put in the form,
\begin{equation}
P=6\tilde{\beta}H^{4}\left(2H^{2}+3\dot{H}\right)    
\end{equation}
Here we see that the above invariant $P$ only consists of first order derivatives and hence the FLRW equations will contain derivatives upto the second order only. The Friedmann equations (\ref{frw1}) and (\ref{frw2}) can be rewritten in the standard form as,
\begin{equation}\label{frw3}
3H^{2}=\kappa\left(\rho_{m}+\rho_{cubic}+\rho_{rad}\right)=\kappa \rho_{eff}
\end{equation}
\begin{equation}\label{frw4}
3H^{2}+2\dot{H}=-\kappa\left(p_{m}+p_{cubic}\right)
\end{equation}
where 
\begin{equation}\label{rhomod}
\rho_{cubic}=6\alpha \tilde{\beta}H^{6}
\end{equation}

\begin{equation}\label{pmod}
p_{cubic}=-6\alpha \tilde{\beta}H^{4}\left(H^{2}+2\dot{H}\right)
\end{equation}
Above we see that these $\rho_{cubic}$ and $p_{cubic}$ are the contribution of the cubic gravity on energy density and pressure respectively, and these modifications are introduced in the action by the cubic invariant $P$. The conservation equations of the respective components will be,
\begin{equation}\label{cons1}
\dot{\rho}_{cubic}+3H\left(\rho_{cubic}+p_{cubic}\right)=0
\end{equation}

\begin{equation}\label{cons2}
\dot{\rho}_{m}+3H\rho_{m}=0
\end{equation}

\begin{equation}\label{cons3}
\dot{\rho}_{rad}+4H\rho_{rad}=0
\end{equation}
In the above, we have considered matter to be pressure-less dust along with radiation. Solving eqns.(\ref{cons2}) and (\ref{cons3}) we get respectively,
$\rho_{m}=\rho_{m0}\left(1+z\right)^{3}$ and $\rho_{rad}=\rho_{rad0}\left(1+z\right)^{4}$, where $\rho_{m0}$ and $\rho_{rad0}$ represents the present energy densities of matter and radiation respectively. The cosmological redshift will be given by $z=\frac{1}{a(t)}-1$. The effective equation of state will be given by,
\begin{equation}\label{eoseff}
\omega_{eff}=\frac{p_{eff}}{\rho_{eff}}=\frac{p_{cubic}}{\rho_{m}+\rho_{cubic}+\rho_{rad}}
\end{equation}
The effective dark energy equation of state will be given by,
\begin{equation}\label{eosde}
\omega_{DE}=\omega_{cubic}=\frac{p_{DE}}{\rho_{DE}}=\frac{p_{cubic}}{\rho_{cubic}}
\end{equation}
Now we define the density parameters as,~~
$\Omega_{DE}=\Omega_{cubic}=\frac{\kappa \rho_{cubic}}{3H^{2}}$,~~~$\Omega_{m}=\frac{\kappa \rho_{m}}{3H^{2}}$, ~~~$\Omega_{rad}=\frac{\kappa \rho_{rad}}{3H^{2}}$.

The deceleration parameter may be defined as,
\begin{equation}\label{dece}
q=-1-\frac{\dot{H}}{H^{2}}=\frac{1}{2}+\frac{3}{2}\left(\omega_{m}\Omega_{m}+\omega_{cubic}\Omega_{cubic}\right)
\end{equation}

Now we can generalize the action for the cubic gravity (\ref{actioncubic}) by including an arbitrary function of the scalar invariant $P$ as given below,

\begin{equation}\label{actionfp}
\mathcal{S}=\int \sqrt{-g}d^{4}x\left[\frac{R}{2\kappa}+f(P)\right]+\int \left(\mathcal{L}_{m}+\mathcal{L}_{rad}\right)\sqrt{-g}d^{4}x
\end{equation}
where $f(P)$ is the arbitrary function of $P$. Varying the action with respect to the metric we get,
\begin{equation}\label{efe1new}
G_{\mu\nu}=\kappa \left(T_{\mu\nu}+ \tilde{H}_{\mu\nu}+T_{\mu\nu}^{rad}\right)
\end{equation}
where $T_{\mu\nu}$ and $T_{\mu\nu}^{rad}$ are given by eqn.(\ref{emt1}) and
\begin{equation}\label{Hnew}
\tilde{H}_{\mu\nu}=-\frac{2}{\sqrt{-g}}\frac{\delta \left(\sqrt{-g}f(P)\right)}{\delta g^{\mu\nu}}=g_{\mu\nu}f(P)+R^{\alpha \beta \rho}{}_{(\mu}\tilde{K}_{\nu)\rho \alpha \beta}+2\nabla^{\alpha}\nabla^{\beta}\tilde{K}_{\alpha(\mu\nu)\beta}
\end{equation}
In the above expression the tensor $\tilde{K}_{\alpha\beta\mu\nu}$ is given in terms of the tensor $K_{\alpha\beta\mu\nu}$ from eqn.(\ref{Kten}) as,
\begin{equation}
\tilde{K}_{\alpha\beta\mu\nu}=f'(P)K_{\alpha\beta\mu\nu}
\end{equation}
where the primes denote derivative with respect to the argument. Considering FLRW geometry, we get the following two FLRW equations,
\begin{equation}\label{flrwfp1}
3H^{2}=\kappa\left(\rho_{m}+\rho_{f_{P}}+\rho_{rad}\right)
\end{equation}
\begin{equation}\label{flrwfp2}
3H^{2}+2\dot{H}=-\kappa\left(p_{m}+p_{f_{P}}\right)
\end{equation}
where 
\begin{equation}\label{rhofp}
\rho_{f_{P}}=-f(P)-18\tilde{\beta}H^{4}\left(H\partial_{t}-H^{2}-\dot{H}\right)f'(P)
\end{equation}

\begin{equation}\label{pfp}
p_{f_{P}}=f(P)+6\tilde{\beta}H^{3}\left[H\partial_{t}^{2}+2\left(H^{2}+2\dot{H}\right)\partial_{t}-3H^{3}-5H\dot{H}\right]f'(P)
\end{equation}
Here $\partial_{t}=\frac{\partial}{\partial t}$ and $\partial_{t}^{2}=\frac{\partial^{2}}{\partial t^{2}}$. The conservation equation for the modified gravity sector is given by,
\begin{equation}\label{cons4}
\dot{\rho}_{f_{P}}+3H\left(\rho_{f_{P}}+p_{f_{P}}\right)=0
\end{equation}
The conservation equations for the matter and the radiation sectors remain same as the ones given in eqns.(\ref{cons2}) and (\ref{cons3}) respectively. In the above equations for $f(P)$ gravity, if we put $f(P)=\alpha P$ we recover the corresponding equations for the cubic gravity.

\section{Observational data analysis}
Here we would like to perform observational data analysis on our
theoretical models using various observational data. We would also
adopt different statistical tools and techniques for our data
analysis methodology. We would concentrate on methods like
$\chi^2$ minimization technique, Markov chain Monte Carlo random
sampling methods, etc. We would also like to verify the validity
of our theoretical model using machine learning techniques. From here on we will use $\kappa=1$ everywhere. 

Firstly, we build up the theoretical model for the cubic gravity. The first FLRW equation of cubic gravity (\ref{frw1}) can be put in the form,
\begin{equation}\label{hubble1}
H^{2}(z)=-\frac{6^{1/3}\alpha \tilde{\beta}+\left[3\left(1+z\right)^{3}\alpha^{2}\tilde{\beta}^{2}\left(\rho_{m0}+\rho_{rad0}\left(1+z\right)\right)+\sqrt{3\alpha^{3}\tilde{\beta}^{3}\left(3\alpha \tilde{\beta}\left(\rho_{m0}+\rho_{rad0}\left(1+z\right)\right)^{2}\left(1+z\right)^{6}-2\right)}\right]^{2/3}}{6^{2/3}\alpha \tilde{\beta}\left[3\left(1+z\right)^{3}\alpha^{2}\tilde{\beta}^{2}\left(\rho_{m0}+\rho_{rad0}\left(1+z\right)\right)+\sqrt{3\alpha^{3}\tilde{\beta}^{3}\left(3\alpha \tilde{\beta}\left(\rho_{m0}+\rho_{rad0}\left(1+z\right)\right)^{2}\left(1+z\right)^{6}-2\right)}\right]^{1/3}}
\end{equation}
The present time dimensionless density parameters can be defined as,
\begin{equation}\label{densitynow}
\Omega_{m0}=\frac{\rho_{m0}}{3H_{0}^{2}}, ~~~~~~~\Omega_{rad0}=\frac{\rho_{rad0}}{3H_{0}^{2}}
\end{equation}
Using these definitions in the expression for Hubble parameter (\ref{hubble1}) we get,
\begin{equation}\label{hubble2}
H^{2}(z)=-\frac{6^{1/3}\alpha \tilde{\beta}+\left[9H_{0}^{2}\left(1+z\right)^{3}\alpha^{2}\tilde{\beta}^{2}\left(\Omega_{m0}+\Omega_{rad0}\left(1+z\right)\right)+\sqrt{3\alpha^{3}\tilde{\beta}^{3}\left(27H_{0}^{4}\alpha \tilde{\beta}\left(\Omega_{m0}+\Omega_{rad0}\left(1+z\right)\right)^{2}\left(1+z\right)^{6}-2\right)}\right]^{2/3}}{6^{2/3}\alpha \tilde{\beta}\left[9H_{0}^{2}\left(1+z\right)^{3}\alpha^{2}\tilde{\beta}^{2}\left(\Omega_{m0}+\Omega_{rad0}\left(1+z\right)\right)+\sqrt{3\alpha^{3}\tilde{\beta}^{3}\left(27H_{0}^{4}\alpha \tilde{\beta}\left(\Omega_{m0}+\Omega_{rad0}\left(1+z\right)\right)^{2}\left(1+z\right)^{6}-2\right)}\right]^{1/3}}    
\end{equation}
The free parameters appearing in the above model are $H_{0}$, $\Omega_{m0}$, $\Omega_{rad0}$, $\alpha$ and $\tilde{\beta}$. From the recent astronomical data we will fix the parameters $H_{0}=67.66~ km~ sec^{-1}Mpc^{-1}$ (Aghanim et al. 2020), $\Omega_{m0}=0.31$ (Ade et. al. 2016) and $\Omega_{rad0}=8.48\times10^{-5}$ (Chavanis, 2015). So we are left with two free parameters and the corresponding parameter space to be constrained is $(\alpha, \tilde{\beta})$.

Now we consider the theoretical framework for $f(P)$ gravity. To further proceed with the FLRW equations (\ref{flrwfp1}) and (\ref{flrwfp2}) we need to consider some specific models. Looking at the complexity of the theory, we consider the simplest power law model,
\begin{equation}\label{power}
f(P)=f_{0}P^{\gamma}
\end{equation}
where $f_{0}$ and $\gamma$ are constants. Using this in the eqn.(\ref{flrwfp1}) we get,

$$3H^{2}=\frac{6^{\gamma}f_{0}\left[\tilde{\beta}H^{4}\left(2H^{2}+3\dot{H}\right)\right]^{\gamma}\left[H^{4}\left(6\gamma-4\right)-3\left(12\gamma^{2}-17\gamma+4\right)\dot{H}H^{2}-9\left(\gamma-1\right)\left(4\gamma-1\right)\dot{H}^{2}-9\gamma\left(\gamma-1\right)H\ddot{H}\right]}{\left(2H^{2}+3\dot{H}\right)^{2}}$$
\begin{equation}\label{powfrw}
+\left[\rho_{m0}+\left(1+z\right)\rho_{rad0}\right]\left(1+z\right)^{3}
\end{equation}
Now we consider the following relations from cosmography,
\begin{equation}\label{cosmogr}
\dot{H}=-\left(1+q\right)H^{2}~,~~~~~~~~~~~~~~\ddot{H}=\left(j+3q+2\right)H^{3}
\end{equation}
where $q$ is the deceleration parameter defined in eqn.(\ref{dece}) and $j$ is the jerk parameter. Using the above parameters in eqn.(\ref{powfrw}) we get,

$$3H^{2}+\frac{6^{\gamma}f_{0}\left[-\tilde{\beta}\left(1+3q\right)H^{6}\right]^{\gamma}\left[\left(1+3q\right)^{2}-3\gamma\left(6+3j+22q+15q^{2}\right)+9\gamma^{2}\left(2+j+7q+4q^{2}\right)\right]}{\left(1+3q\right)^{2}}$$

\begin{equation}\label{powfrw2}
=\left[\rho_{m0}+\left(1+z\right)\rho_{rad0}\right]\left(1+z\right)^{3}   
\end{equation}

We see that it is quite difficult to get analytical expression for $H(z)$ from the above equation because of its highly non-linear nature. Fortunately for $\gamma=-1/3$ we get an analytic expression for $H(z)$ as given below,

$$H^{2}(z)=\frac{1}{2}H_{0}^{2}\left[\Omega_{m0}+\Omega_{rad0}\left(1+z\right)\right]\left(1+z\right)^{3}-\frac{1}{36}\left[2\times 6^{2/3}\left(\frac{f_{0}^{3}\left(9+4j+35q+28q^{2}\right)^{3}}{\left(1+3q\right)^{7}\tilde{\beta}}\right)^{1/3}\right.$$

\begin{equation}\label{powfr3}
\left.+9H_{0}^{4}\left[\Omega_{m0}+\Omega_{rad0}\left(1+z\right)\right]^{2}\left(1+z\right)^{6}\right]^{1/2}
\end{equation}

This will serve as our theoretical tool in this analysis. The free parameters appearing in the above model are $H_{0}$, $\Omega_{m0}$, $\Omega_{rad0}$, $f_{0}$, $\tilde{\beta}$, $q$ and $j$. Just like cubic gravity, we will fix the parameters $H_{0}=67.66~ km~ sec^{-1}Mpc^{-1}$, $\Omega_{m0}=0.31$ and $\Omega_{rad0}=8.48\times10^{-5}$. In addition these we will also fix the deceleration $q=-0.503$ (Capozziello, D'Agostino \& Luongo 2019; Aghanim et al. 2020). From the combination of three kinematical datasets: the gold sample of type Ia supernovae (Riess et al. 2004), the SNIa data from the SNLS project (Astier et al. 2006) and the X-ray galaxy cluster distance measurements (Rapetti et al. 2007) the value of the jerk parameter is estimated as $j=2.16^{+0.81}_{-0.75}$. So for the jerk parameter we will use the value $j=2.16$ for this study. So in this model, we are left with two free parameters and the corresponding parameter space to be constrained is $(f_{0}, \tilde{\beta})$.

\subsection{Analysis with Cosmic Chronometer (CC) Data}
In this work we will use the $z-H(z)$ cosmic chronometer data sets (Jimenez \& Loeb 2002; Simon, Verde \& Jimenez 2005; Stern et al. 2010; Zhang et al. 2014; Moresco 2015). The complete data table for CC data can be found in the refs. Rudra \& Giri (2021) and Ranjit, Rudra \& Kundu (2021). We observe that the resdshift range of the data is $(0, 2)$. Considering the current redshift to be $z\approx0$ and the universe is evolving from higher redhift regime to a lower redshift regime, the span of the data seems to be quite relevant cosmologically. The cosmic choronometers are a very useful set of tools in understanding the gradual evolution of the universe. This data is extracted from the observation of passively evolving primordial galaxies, using the technique of differential age evolution. We know that for a universe modelled by the FLRW equations the relation between the Hubble parameter $H$ and the redshift $z$ can be given by $H=-\left(1+z\right)^{-1}dz/dt$. So the time gradient of the redshift $z$ is measured and used in the above relation to find the Hubble parameter values. The complete data set of the cosmic chronometers spans around 10 Gyr of cosmic time.

We would like to perform a data analysis with the 30 point CC data-set and constrain the free parameters of the models. To achieve this, we will first establish the $\chi^{2}$ statistic as a sum of standard normal variate as follows:
\begin{equation}\label{chicc}
{\chi}_{CC}^{2}=\sum\frac{\left[H(z)-H_{obs}(z)\right]^{2}}{\sigma^{2}(z)}
\end{equation}
where $H(z)$ and $H_{obs}(z)$ are the theoretical and
observational values of Hubble parameter at different red-shifts respectively and $\sigma(z)$ is the standard deviation representing the corresponding error in measurement of the data point. Here we consider the present value of Hubble parameter as $H_{0}$ = 69 $\pm$ 8 Km s$^{-1}$
Mpc$^{-1}$ and we also consider a fixed prior distribution for it. The reduced chi square value can be given as
\begin{equation}\label{reducechicc}
L=\chi_{R}^{2}= \int e^{-\frac{1}{2}{\chi}_{CC}^{2}}P(H_{0})dH_{0}
\end{equation}
where $P(H_{0})$ is the prior distribution function for $H_{0}$.

\subsection{Joint analysis with CC+BAO}
In the work of Eisenstein et al. (2005), we found an elegant method of clubbing the cosmological data with the Baryon acoustic oscillation (BAO) peak parameter to refine the constraining methods of model parameters.  The BAO signal was detected for the first time while conducting the Sloan digital sky survey (SDSS) at around a scale of 100 Mpc. Another success of the SDSS survey was that, it confirmed the observations and results obtained from the Wilkinson microwave anisotropy probe (WMAP). During the SDSS survey spectroscopic samples were retrieved from thousands of red galaxies that spanned across the sky, covering a diameter of around five billion light yeras. The BAO peak parameter is defined as (Thakur, Ghose \& Paul 2009; Paul, Thakur \& Ghose 2010; Paul, Ghose \& Thakur 2011; Ghose, Thakur \& Paul 2012):
\begin{equation}\label{baopeak}
{\cal A}=\frac{\sqrt{\Omega_{m}}}{E(z_{1})^{1/3}}
\left(\frac{1}{z_{1}}~\int_{0}^{z_{1}}
\frac{dz}{E(z)}\right)^{2/3}
\end{equation}
In the above expression $E(z)=H(z)/H_{0}$ is called the normalized Hubble parameter. From the SDSS survey it is well known that  $z_{1}=0.35$ is the prototypical value of red-shift which we will use in our analysis. For a flat model of universe SDSS data estimates the value of the peak parameter to around ${\cal A}=0.469\pm 0.017$ (Eisenstein et al. (2005)). The $\chi^{2}$ function for the BAO measurement can be given as

\begin{equation}\label{Eqn1.5.174}
\chi^{2}_{BAO}=\frac{({\cal A}-0.469)^{2}}{(0.017)^{2}}
\end{equation}
The cumulative analysis with cosmic chronometers and the BAO peak parameter (CC+BAO) for the $\chi^{2}$
function can be defined as (Wu \& Yu 2007; Thakur, Ghose \& Paul 2009; Paul, Thakur \& Ghose 2010; Paul, Ghose \& Thakur 2011; Ghose, Thakur \& Paul 2012)
\begin{equation}\label{Eqn1.5.175}
\chi^{2}_{total}=\chi^{2}_{CC}+\chi^{2}_{BAO}
\end{equation}
The above given cumulative $\chi^{2}$ will be used in our analysis to refine the results obtained for CC data by the additional restrictions of the BAO peak parameter.

\subsection{Joint analysis with CC+BAO+CMB}
With the discovery of the late time cosmic acceleration general relativity became inconsistent at the cosmological scales. To explain such a phenomenon the scientific community has resorted to the concept of an exotic fluid with negative pressure known as dark energy. Cosmic microwave background (CMB) observations is the most interesting probe that provides some observational evidence to the theoretical framework of dark energy via its power spectrum. One disadvantage of the CMB parameter is that it is not sensitive to the background perturbations. However the parameter is perfectly suitable to constrain cosmological models using its peak.
The first peak of the CMB power spectrum is primarily a shift parameter given by
\begin{equation}\label{Eqn1.5.176}
{\cal R}=\sqrt{\Omega_{m}} \int_{0}^{z_{2}} \frac{dz}{E(z)}
\end{equation}
where $z_{2}$ is the value of redshift consistent with the last scattering surface. From the Wilkinson microwave anisotropy probe (WMAP) 7-year data, which is available in the work of Komatsu et al. (2011) the value of the above shift parameter has been obtained 
as ${\cal R}=1.726\pm 0.018$ at the redshift $z=1091.3$. Now the $\chi^{2}$ function for the CMB measurement can be given in terms of the CMB shift parameter by the following relation,
\begin{equation}\label{Eqn1.5.177}
\chi^{2}_{CMB}=\frac{({\cal R}-1.726)^{2}}{(0.018)^{2}}
\end{equation}
Now considering the CC data constrained with both the BAO and CMB peaks together, we can perform the joint data analysis for CC+BAO+CMB. The total $\chi^{2}$ function for this case can be given as,
\begin{equation}\label{Eqn1.5.178}
\chi^{2}_{TOTAL}=\chi^{2}_{CC}+\chi^{2}_{BAO}+\chi^{2}_{CMB}
\end{equation}
Using this we can refine the results obtained for the CC+BAO case, and put better constraints on the free parameters of the models. Moreover since CMB observations are motivated from the dark energy, they serve as the best tools to constrain modified gravity theories when compiled with other data sets.

\section{Statistical Analysis and Machine Learning}
In this section, we will briefly describe the statistical methods and the supervised machine learning techniques which will be used in view of the statistical and machine learning perspective. Three different methods will be studied namely, analysis with correlation coefficients, linear regression analysis and artificial neural network. We discuss the methods below:

\subsection{Correlation Coefficients}
Correlation coefficients (here after $R$) (Heumann, Schomaker \& Shalabh 2016) are used to measure the strength of the linear relationship between two variables. $R$ is a statistical technique used to determine the degree to which two variables are related (i.e. for a bi-variate sample). 
In this work, there are two data sets ${\bf z}$ and $\bf H$ of length $N$, where $N = 30$. Hence, ${\bf z}  = ({z_1, z_2,z_3,\dots,z_N})$ and ${\bf H}  = ({H_1, H_2,H_3,\dots,H_N})$. Then,  $R$ is given by 
\begin{equation}\label{Eqn1.5.176new}
R = 
\frac{
N\sum_{i=1}^{N}{{z_i}{H_i}}-\left(\sum_{i=1}^{N}{z_i}\sum_{n=1}^{N}{H_i}\right)}
{
\sqrt{
\left[N \sum_{i=1}^{N}{{z_i}^2}-\left(\sum_{i=1}^{N}{z_i}\right)^2 \right]
\left[N \sum_{i=1}^{N}{{H_i}^2}-\left(\sum_{i=1}^{N}{H_i}\right)^2 \right]
}
}
\end{equation}
The possible range of values for the correlation coefficient is $-1.0$ to $1.0$. If the correlation coefficient is greater than zero, it is a positive relationship. Conversely, if the value is less than zero, it is a negative relationship. A value of zero indicates that there is no relationship between the two variables.If $R$ is close to $1$ or $-1$ there is a strong positive or negative correlation respectively.

\subsection{Linear Regression (LR)}
The regression analysis is a common but powerful supervised machine learning algorithm, which helps to predict the trends and future values from the existing data. In the context of regression models, the simple linear regression (LR) (Sen \& Srivastava 1990) is one of the most basic and common predictive analysis model. Basically, LR is used to predict the relationship between independent (known as input/s) and dependent variables (known as output) assuming a linear relationship between those input/s and output. If there is a single input variable, then the model is referred to as LR, while if there are multiple input variables, the same is termed as multiple linear regression model (MLR). In both LR and MLR, the output will be a single variable. The distribution of regression residuals are normal distribution. In this work we have only one input parameter,viz., $z$ and output parameter $H_{LR}$. So, we used LR here.If 
\begin{equation}\label{Eqn1.5.177new}
    \bar{z} = \sum_{i=1}^{N}{z},
\end{equation}
and 
\begin{equation}\label{Eqn1.5.178new}
  \bar{H} = \sum_{i=1}^{N}{H_i},  
\end{equation}

then, the estimated value of $H_{LR}$ is obtained from the equation of straight 
line for LR which is given by
\begin{equation}\label{Eqn1.5.179}
{H}_{LR} =  \alpha z + \beta
\end{equation}
where, the slope of the straight line is given as,
\begin{equation}\label{Eqn1.5.180}
\alpha = \frac{\sum_{i=1}^{N}(z_i - \bar{z}) (H_i - \bar{H})} {\sum_{i=1}^{N}(z_i - \bar{H})^2},
\end{equation}
and the intercept of LR line is,
\begin{equation}\label{Eqn1.5.181}
\beta = \bar{H} - {\alpha \bar{z}}.
\end{equation}

\subsection{Artificial Neural Network (ANN)}
An artificial neuron network is a computational model that mimics the way nerve cells work in the human brain (MacGregor \& Lewis 1977).  Human brain consists of many neurons connected to each of their neighbours. In human brain, each neurons pass the input signal from one to another as well as pass the information that is to be computed for output. Similarly, the ANN also pass the input signal from one neuron to another and create a network of artificial neurons for computation. A simple ANN structure consists one or more number of inputs and a single output. Let, the net input to a process is $H_{in}$. Then the net output $H_{out}$ is a function of $H_{in}$. For a simple ANN the net output is considered as a binary step function as below. 

\begin{equation}\label{Eqn1.5.182}
H_{out} = F_{ann}(H_{in}) =
  \begin{cases}
    			1 & \text{if $H_{in} > 0$} \\   
    			0 & \text{if $H_{in} \leq 0$}           
  \end{cases}
\end{equation}

All the neurons in ANN build the layers or network by interconnecting  themselves. These interconnection may or may not be fully connected. According to this layered architecture the ANN can be classified into various divisions,viz, single layer feed forward ANN, multi-layer feed forward ANN, competitive network and recurrent network. All of these networks may have one or more hidden layer with the input and output layers. The output only generates from an output processing unit when a special function satisfies required criteria for given input variables. This special function that maps the net input value to the output values is known as activation function of that output unit of the ANN. It uses learning algorithms that can independently make adjustments - or learn, in a sense - as they receive new input. This makes them a very effective tool for non-linear statistical data modeling.

\section{Results}
 In connection with this work, we have developed a complete PYTHON package, where, using any theoretical model for $H(z)$, we can find the best fitted auxiliary parameters, bounds of free parameters, plotting of parameter space, statistical analysis and estimation analysis for $H$ using machine learning. Although we have developed our own code almost, however, in order to find bounds of free parameters at different statistical confidence intervals, we took the help of the publicly available \textit{CosmoMC} code (Lewis, Challinor \& Lasenby 2000; Lewis \& Bridle 2002). We found various interesting results which will be presented in the following subsections.

\subsection{Auxiliary Parameter Analysis}
It is relevant to mention here that these model have some background theoretical motivations and hence the results for this model may be interesting for referencing other results. After fixing the known parameters, we are left with only two free parameters,viz., ($\alpha$, $\tilde{\beta}$) for Cubic gravity and ($f0$, $\tilde{\beta}$) for $f(P)$ gravity. So computationally this seems to be a relatively convenient scenario. Using our code we have constrained the theoretical models with the data and put bounds on the free parameters. The results are given in the tables \ref{table:T2A} and \ref{table:T2B}.  
\begin{table}[h!]
\centering
\begin{tabular}{l@{\hskip .1in}c@{\hskip .5in} c@{\hskip .5in}  r}
\hline
Data   &  $\alpha$ & $\tilde{\beta}$  & ${\chi^{2}_{min}}$ \\
\hline
CC & $-1.3 \times 10^{-6}$ & $2.9 \times 10^{-7}$  & $0.4256$\\
CC+BAO & $-7.4 \times 10^{-6}$ & $8.7 \times 10^{-7}$  & $17.3336$\\
CC+BAO+CMB & $-7.27 \times 10^{-6}$ & $7.56 \times 10^{-7}$  & $49.7404$\\
\hline
Data   &  $f0$ & $\tilde{\beta}$  & ${\chi^{2}_{min}}$ \\
\hline
CC & $-3.769 \times 10^{-4}$ & $0.2385$  & $0.5469$\\
CC+BAO & $-4.423 \times 10^{-4}$ & $0.2354$  & $22.4963$\\
CC+BAO+CMB & $-4.1635 \times 10^{-4}$ & $0.2393$  & $51.7947$\\
\hline
\end{tabular}
\vspace{3mm}
\centering \caption{The best fit values of $\alpha$ and $\tilde{\beta}$ for Cubic gravity, with the minimum values of $\chi^2$ presented in the top pane. In the lower pane we have the best fit values for $f_{0}$ and $\tilde{\beta}$ for $f(P)$ gravity with the minimum values of $\chi^2$.}
\label{table:T2A}
\end{table}

\begin{table}[h!]
\centering
\begin{tabular} {l  c@{\hskip .3in} c@{\hskip .3in} c@{\hskip .3in} r}
\hline
Parameter &~~ 68\% limits & 95\% limits & 99\% limits\\
\hline
\\ $\alpha$  &  $-4.3 \times 10^{{-6}^{+1.8 \times 10^{-6}}}_{-1.8 \times 10^{-6}}$ 
&  $-4.3 \times 10^{{-6}^{+2.9 \times 10^{-6}}}_{-2.9 \times 10^{-6}}$ 
&  $-4.3 \times 10^{{-6}^{+3.0 \times 10^{-6}}}_{-3.0 \times 10^{-6}}$  \\ \\
$\tilde{\beta}$  &  $5.8 \times 10^{{-7}^{+1.7 \times 10^{-7}}}_{-1.7 \times 10^{-7}}$ 
& $5.8 \times 10^{{-7}^{+2.8 \times 10^{-7}}}_{-2.8 \times 10^{-7}}$ 
&  $5.8 \times 10^{{-7}^{+2.9 \times 10^{-7}}}_{-2.9 \times 10^{-7}}$    \\ \\
\hline 
\\ $f0$  & $-6.9 \times 10^{{-4}^{+1.8 \times 10^{-4}}}_{-1.8 \times 10^{-4}}$  
&  $-6.9 \times 10^{{-4}^{+3.0 \times 10^{-4}}}_{-3.0 \times 10^{-4}}$ 
&  $-6.9 \times 10^{{-4}^{+3.1 \times 10^{-4}}}_{-3.1 \times 10^{-4}}$  \\ \\
$\tilde{\beta}$  &  $0.169^{+0.040}_{-0.040}$ 
& $0.169^{+0.066}_{-0.0}$ 
&  $0.169^{+0.069}_{-0.069}$     \\ 
\\ \hline
\end{tabular}
\vspace{3mm}
\centering \caption{Bounds on the free parameters for Cubic (top) and $f(P)$ (bottom) gravity from CC data for different confidence limits}
\label{table:T2B}
\end{table}

\begin{figure}[hbt!]
\centering
\includegraphics{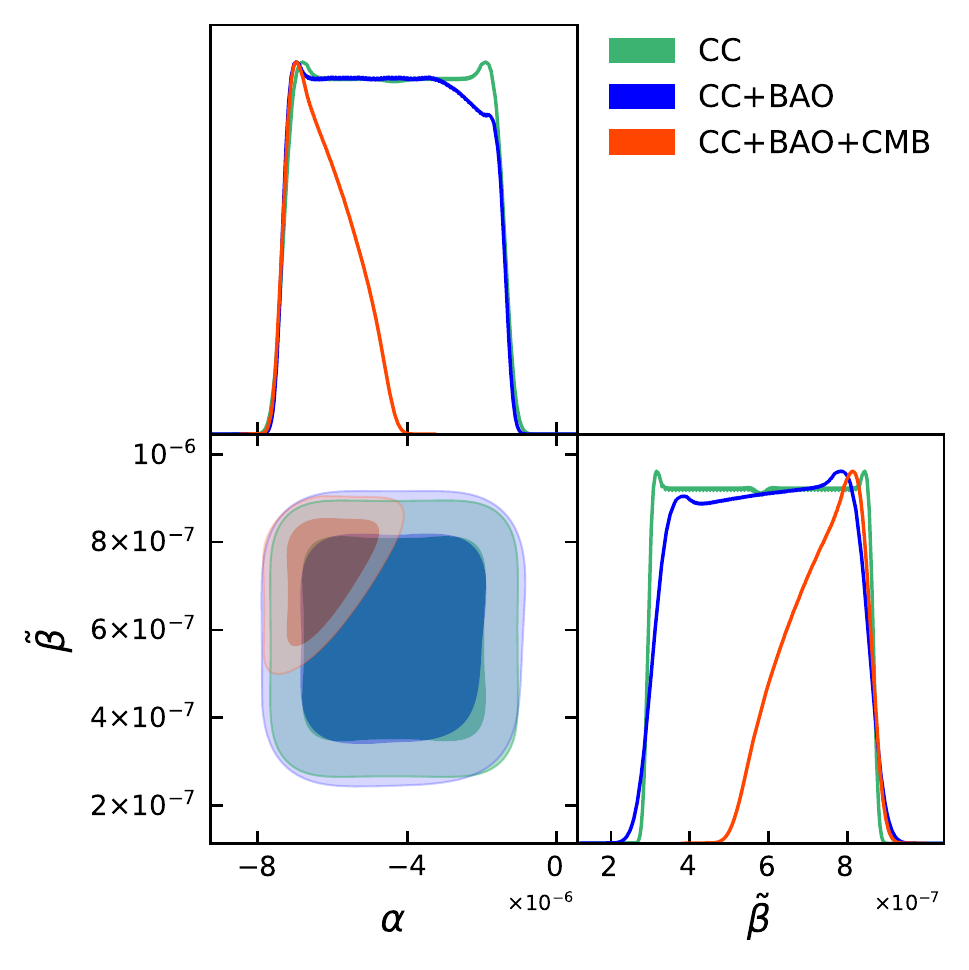}
\centering \caption{1D distributions and 2D joint likelihood contours of the free parameters ($\alpha, \tilde{\beta}$) of cubic gravity model. The deeper shades show the $68\%$ confidence intervals and the lighter shades represent the $95\%$ confidence intervals for the parameters.}
\label{Fig1a}
\end{figure}

\begin{figure}[hbt!]
\centering
\includegraphics{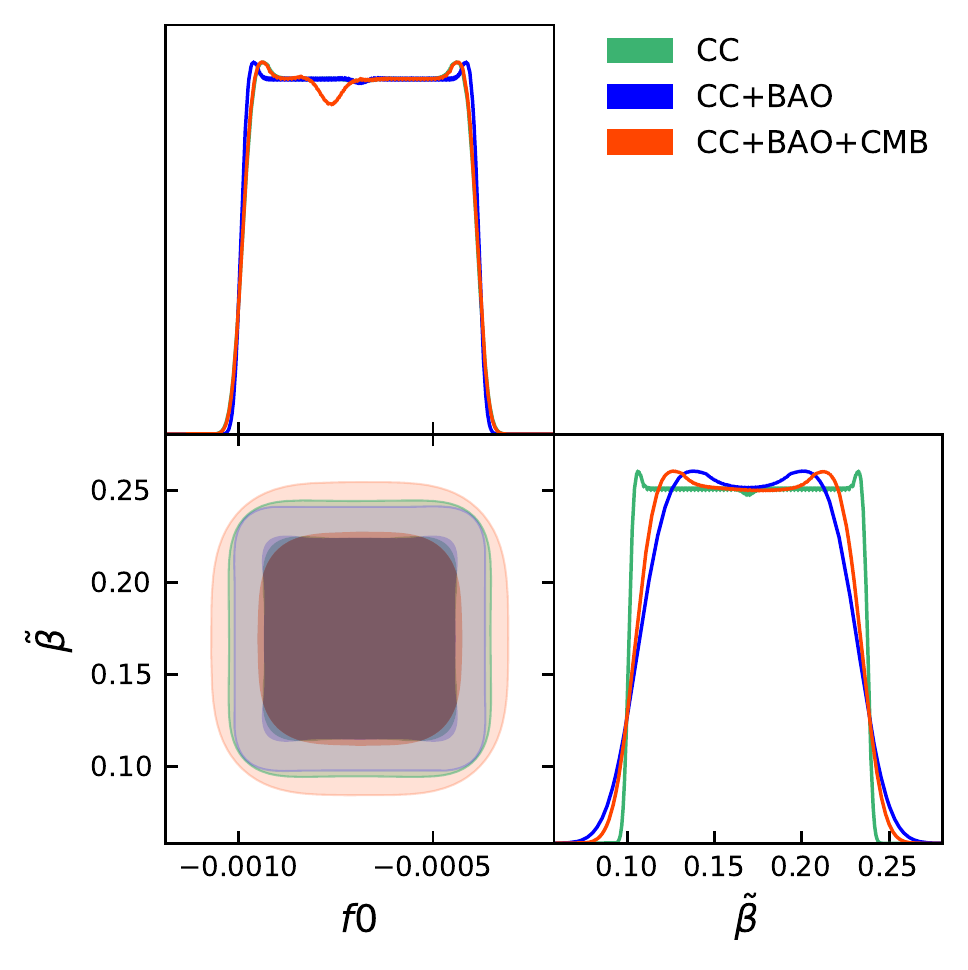}
\centering \caption{1D distributions and 2D joint likelihood contours of the free parameters ($f0$, $\tilde{\beta}$) of $f(P)$ gravity model. The deeper shades show the $68\%$ confidence intervals and the lighter shades represent the $95\%$ confidence intervals for the parameters.}
\label{Fig1b}
\end{figure}

Fig. \ref{Fig1a} and Fig. \ref{Fig1b} shows the 2D confidence contours and the distributions followed by the free parameters of cubic and $f(P)$ gravities respectively. From distribution curves we see that for almost all parameters, we get a Gaussian like distribution (not perfectly Gaussian) for any data-set. For cubic gravity in Fig. \ref{Fig1a} we see that the distribution curves of $\alpha$ for CC and CC+BAO datasets have a flat top with centre at around $-4$, and almost coincide with each other. But for CC+BAO+CMB the distribution is considerably skewed towards left in comparison to the others. For the distributions of $\tilde{\beta}$ the centre shifts to $6$ and the curve for CC+BAO+CMB is skewed towards the right in comparison to the other curves. In Fig.\ref{Fig1b} we have the results for $f(P)$ gravity where it is observed that the distribution curves for the parameters $f_{0}$ and $\tilde{\beta}$ are almost coincident for all the three datasets. The impact of the BAO and CMB peak parameters on the CC data seem to be quite negligible. Almost all the distribution curves have a flat top holding to nearly similar values for a long range. This gives a steady picture of the free parameters.

\subsection{Correlation Analysis}
We will now investigate the corresponding correlation between
observed values of $H$,viz., $H_{obs}$ and the theoretical values of the same $H_{theo}$ for $CC$ data 
with the help of correlation coefficient $R$ (\ref{Eqn1.5.176}). From, Table \ref{table:T2A}, 
it is evident that the best fitting values of the auxiliary parameters; 
$\alpha$ and $\tilde{\beta}$ are $-1.3 \times 10^{-6}$ and $2.9 \times 10^{-7}$ respectively  for cubic gravity model, while, $f0$ and $\tilde{\beta}$ are $-3.769 \times 10^{-4}$ and $0.2385$ respectively for $f(P)$ gravity.  Next, for both of these models, with these specific $\alpha$, $f0$ and $\tilde{\beta}$, we calculated the respective values  of $H_{theo}(z)$ using \ref{hubble2} and \ref{powfr3}. 
In table. \ref{MT}, the calculated values of $H_{theo}$  for cubic gravity model
are given in the third column, while, in the column third column of table \ref{MT1} similar values obtained from $f(P)$ model are provided. In both the tables, the respective observed values are kept in the second column for meaningful comparison. However, using (\ref{Eqn1.5.176}), we calculate that $R$ = $0.9507$ and $0.9498$ for cubic and $f(P)$ model respectively. Both the values of $R$ clearly indicate that, there is a strong positive co-relationship between $H_{obs}$ and our deduced $H_{theo}$ for both cubic and $f(P)$ theories. 

\begin{figure}[hbt!]
\centering
\includegraphics{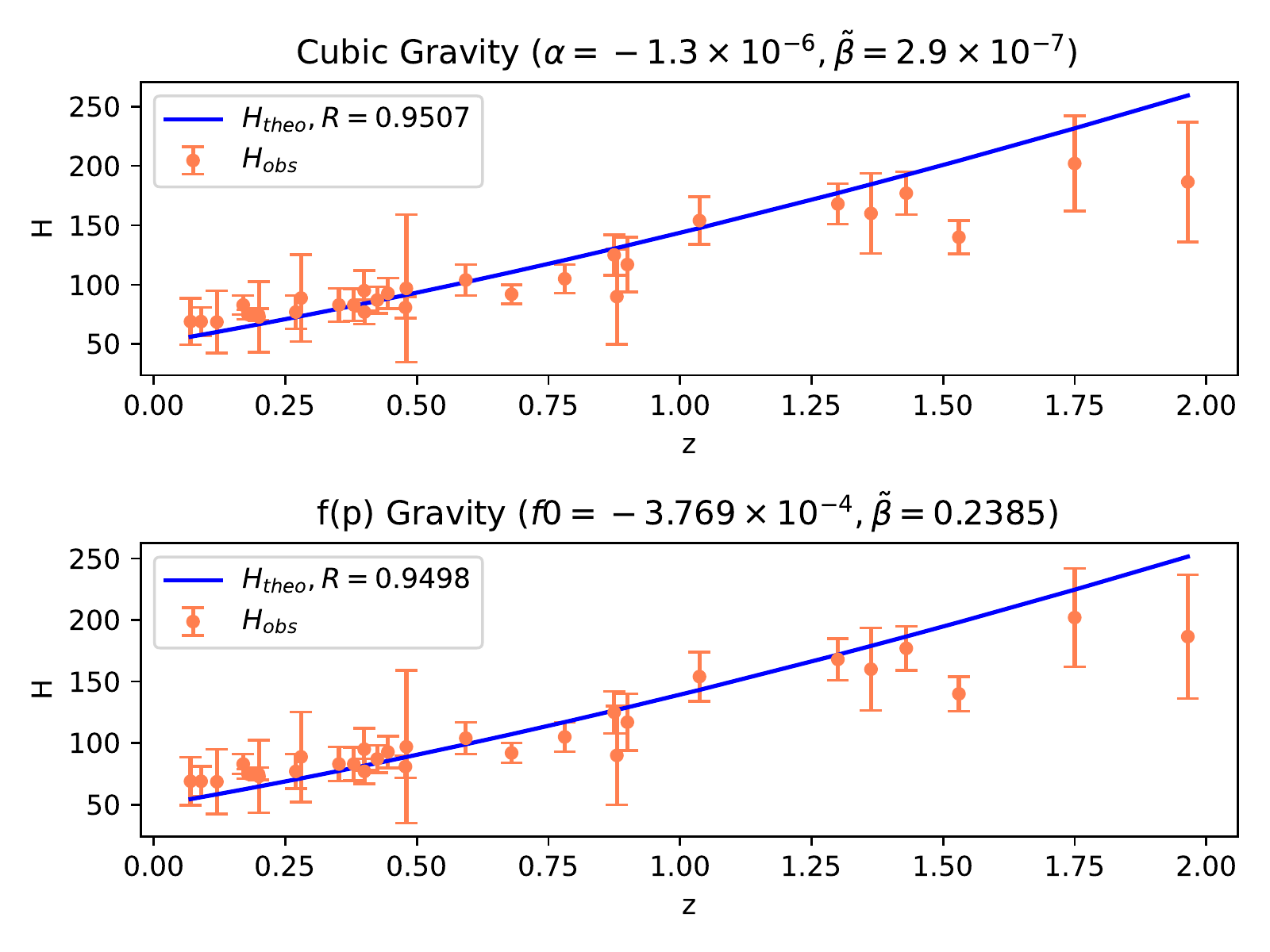}
\caption{$H-z$ relations for observational and theoretical (cubic gravity (top panel) and $f(p)$ gravity (bottom panel) aspects.}
\label{Fig2}
\end{figure}

In Fig.\ref{Fig2} using $H_{obs}$ and $H_{theo}$ values, we have compared our theoretical results for both cubic and $f(P)$ theories with observations in context of $R$.
The top panel of Fig.\ref{Fig2} shows the $H-z$ variations for cubic theory while, the bottom panel represents the same for $f(P)$ theory. In both the cases, best fitted parameters obtained from table \ref{table:T2A} are used to show the $z$ variation of both theoretical and observational $H$. It is interesting to mention here that for cubic gravity model, we got three important observations as follows: (a) $\alpha$ and $\tilde{\beta}$ must be of opposite signs for best fitting values (b) Both of these values of best fitting parameters tends to zero in the context of providing minimum ${\chi}^2$. (c) $H_{obs}$ and $H_{theo}$ data sets are in good agreement with each other. In contrast to this, for the $f(P)$ theory, we found that (a) $f0$ and $\tilde{\beta}$ must be of opposite signs for best fitting values and (b) $H_{obs}$ and $H_{theo}$ data sets are in good agreement with each other. From this it is evident that our fitting analysis has good degree of precision and the constraints on the parameters are excellent.

\subsection{Machine Learning Analysis}
There are several supervised machine learning techniques. Few of these are Linear Regression (LR), Artificial Neural Network (ANN), Supporting Vector Machine (SVM), Random Forest etc.  As mentioned earlier, here, we have used two such algorithms, viz., LR and ANN for the same purpose. This is relevant to mention here that standard observational likelihood analysis (SOLA) is used get more robust parameter estimates but it can't forecast or predict the H(z) with it's known values. Hence the purposes of SOLA and ML are totally different although both are statistical methods. By introducing machine learning techniques we complement our standard observational likelihood analysis, with a more refined form of study. It is known that ML methods are designed to produce far better and accurate predictions than standard statistical procedures. So the motivation behind introducing ML in our study is simply to get better predictions from our theoretical model and the observational data. Moreover we can also compare our results with those obtained from the standard likelihood analysis and get an idea about the range of deviation in the two procedures.
In this part of our work, we have focused on the theoretical values of $H$ (third column of both the Tables \ref{MT} and \ref{MT1}) to investigate the role of machine learning in order to let the computer learn how the function $H(z)$ depends on the red shift parameter $z$. It is well known that for applying any supervised learning techniques, the input data set is segregated in to two subsets. In general, among these two sub sets,the larger set contains more than 50 \% of the data 
which is termed as the training data, while the rest of the set 
contains less than 50\% of data which is known as the testing data. Here we have used the input data as z and target output as H(z) (obtained from our model). We used ~67 \% of this data (randomly taken) for the training purpose for both of the cases of LR and ANN, while the remaining ~33 \% data were used for the testing purpose. For the validation purpose, we compared these machine learning predicted H(z) with both the original H(z) (from $33 \%$ test data) and also with observational H(z). Now, if we include a new set of z values, we don't need to run our theoretical model once again, we can easily use these machine learning techniques to predict the new H(z) corresponding to these new z.
In our present context, we let the machine select randomly the data from $20$ rows given in the
second and third column of Table \ref{MT} and Table \ref{MT1} and estimate the data in the remaining $10$ rows after
learning the evolution of $H(z)$. Here, we discuss the results obtained from LR and ANN methods and also compare them with the observational data sets and theoretical models.

\subsubsection{LR Analysis}
\begin{figure}[hbt!]
\centering
\includegraphics{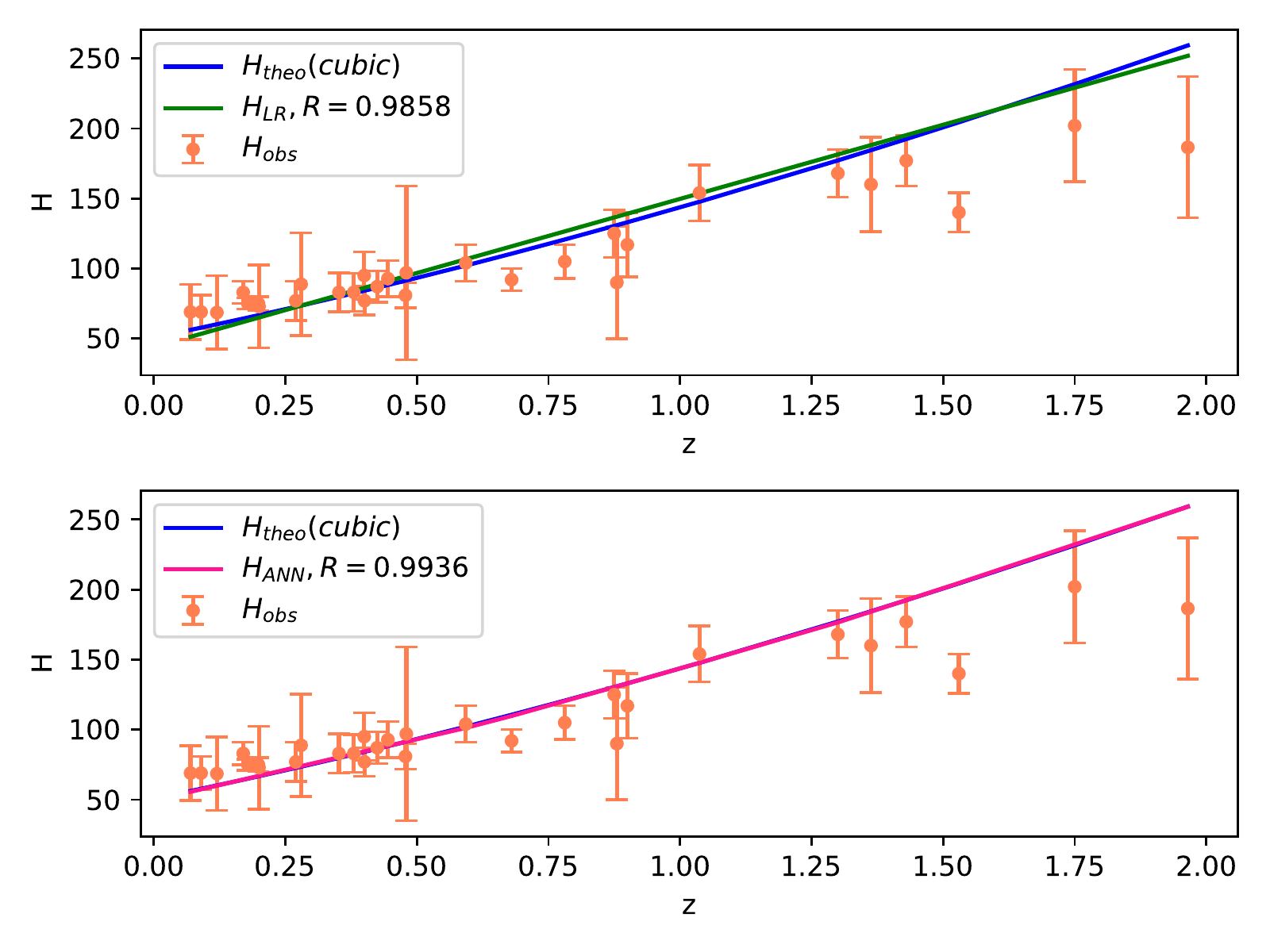}
\centering\caption{For Cubic model: Plots of the estimated values of $H_{LR}(z)$ (top) and $H_{ANN}(z)$ (bottom) with the help of best fitted $H_{theo}$ and the observational $H_{obs}$ with error parameters.}
\label{Fig3a}
\end{figure}

 The LR estimated values of $H$ (denoted as $H_{LR}$) are shown in column 4 of Table \ref{MT} (for cubic gravity model) and Table \ref{MT1} (for $f(P)$ gravity model). Upper panel of Fig.\ref{Fig3a} and Fig.\ref{Fig4a} show the correlation coefficient $R$ of the predicted $H_{LR}$ and our theoretical values $H_{theo}$ for cubic model is $0.9858$, while, same for the $f(P)$ model is almost equal to $0.9838$. As both the values of $R$ for our models are very much close to $1$ and also from the plots (upper panels of Fig. \ref{Fig3a} and Fig. \ref{Fig4a}), we can easily conclude that the prediction using LR is significantly good for the theoretical values obtained from our derived expressions in eqns.\ref{hubble2} and \ref{powfr3} . 

\begin{figure}[hbt!]
\centering
\includegraphics{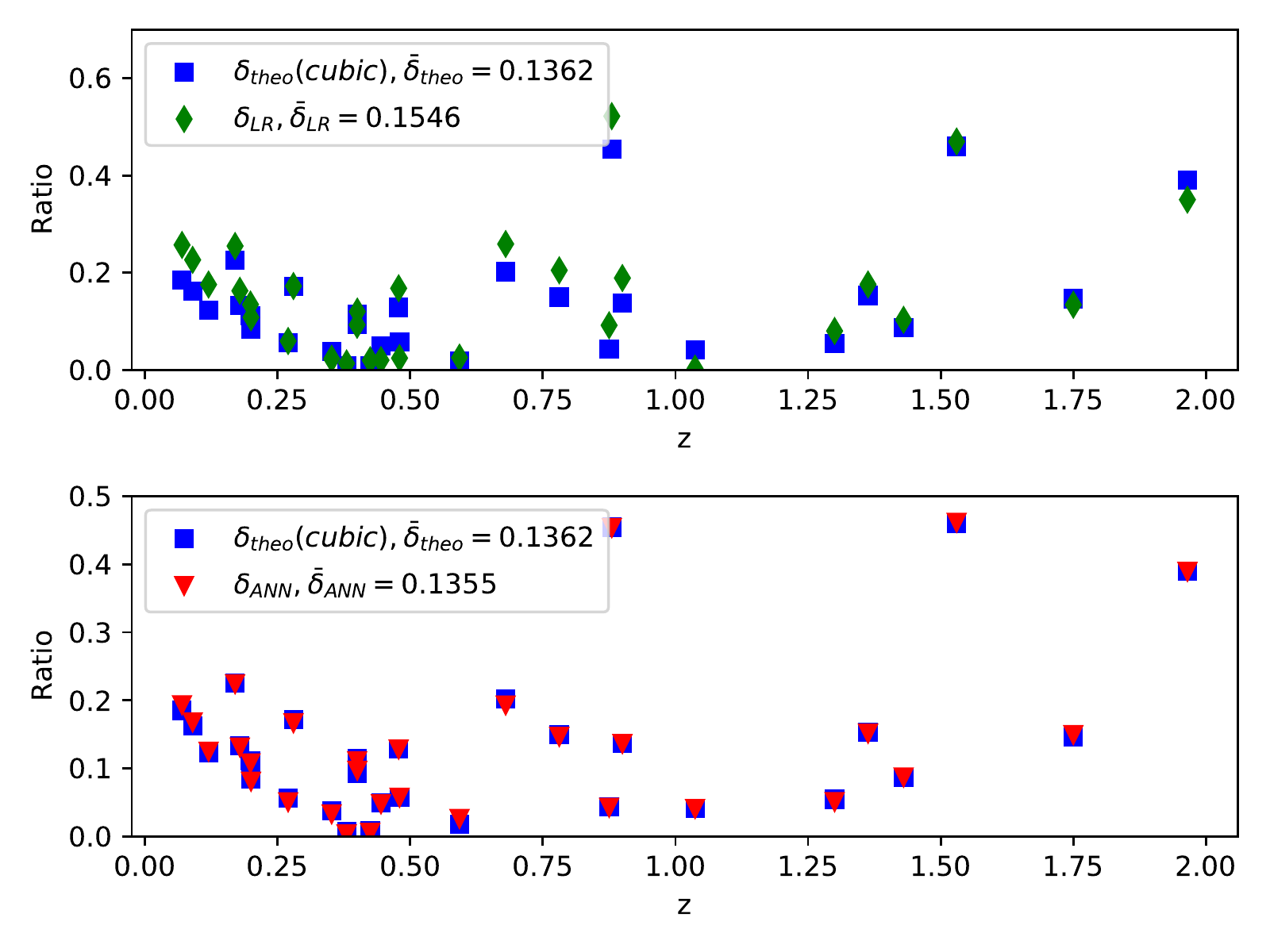}
\centering\caption{For Cubic Model: Comparing the deviation parameters ${\delta}_{theo}$ and ${\delta}_{LR}$ (top) and ${\delta}_{theo}$ and ${\delta}_{ANN}$ (bottom) with the help of best fitting values of the free parameters.}
\label{Fig3b}
\end{figure}

In order to compare $H_{obs}$ with $H_{theo}$ and $H_{LR}$, we provide the ratios 
$H_{theo}/H_{obs}$ and $H_{LR}/H_{obs}$ in column 6 and column 7 of Table \ref{MT} (for cubic gravity theory) and in Table \ref{MT1} (for $f(P)$ theory).  We have also calculated the deviation parameters (Kangal, Salti \& Aydogdu 2019)
\begin{equation}\label{Eqn1.5.182new}
 {\delta}_{theo} = |\frac{H_{theo}}{H_{obs}} -1|,
\end{equation}
and
\begin{equation}\label{Eqn1.5.183}
 {\delta}_{LR} = |\frac{H_{LR}}{H_{obs}} - 1|.   
\end{equation}
We presented these two deviations in the columns 9 and 10 of table \ref{MT} and table \ref{MT1} for the respective models.  The mean values of the deviation percentages are:
For cubic gravity model they are
\begin{equation}\label{Eqn1.5.184}
(\bar{\delta}_{theo}, \bar{\delta}_{LR}) = (0.1362, 0.1546),
\end{equation}
while, for $f(P)$ gravity model we have
\begin{equation}\label{Eqn1.5.184a}
(\bar{\delta}_{theo}, \bar{\delta}_{LR}) = (0.1362, 0.1495).
\end{equation}
In the upper panels of Fig \ref{Fig3b} and \ref{Fig4b} , we plot the deviations ${\delta}_{theo}$ and ${\delta}_{LR}$ along $z$ for respective models.  
One can see from these above mentioned figures that the 
LR values of the deviations for both the models roughly merged with the deviations of the theoretical values. Hence, we can say that the theoretical framework of both the gravity models are well-built and suitable to make interesting cosmological predictions. 


\subsubsection{ANN Analysis}
\begin{figure}[hbt!]
\centering
\includegraphics{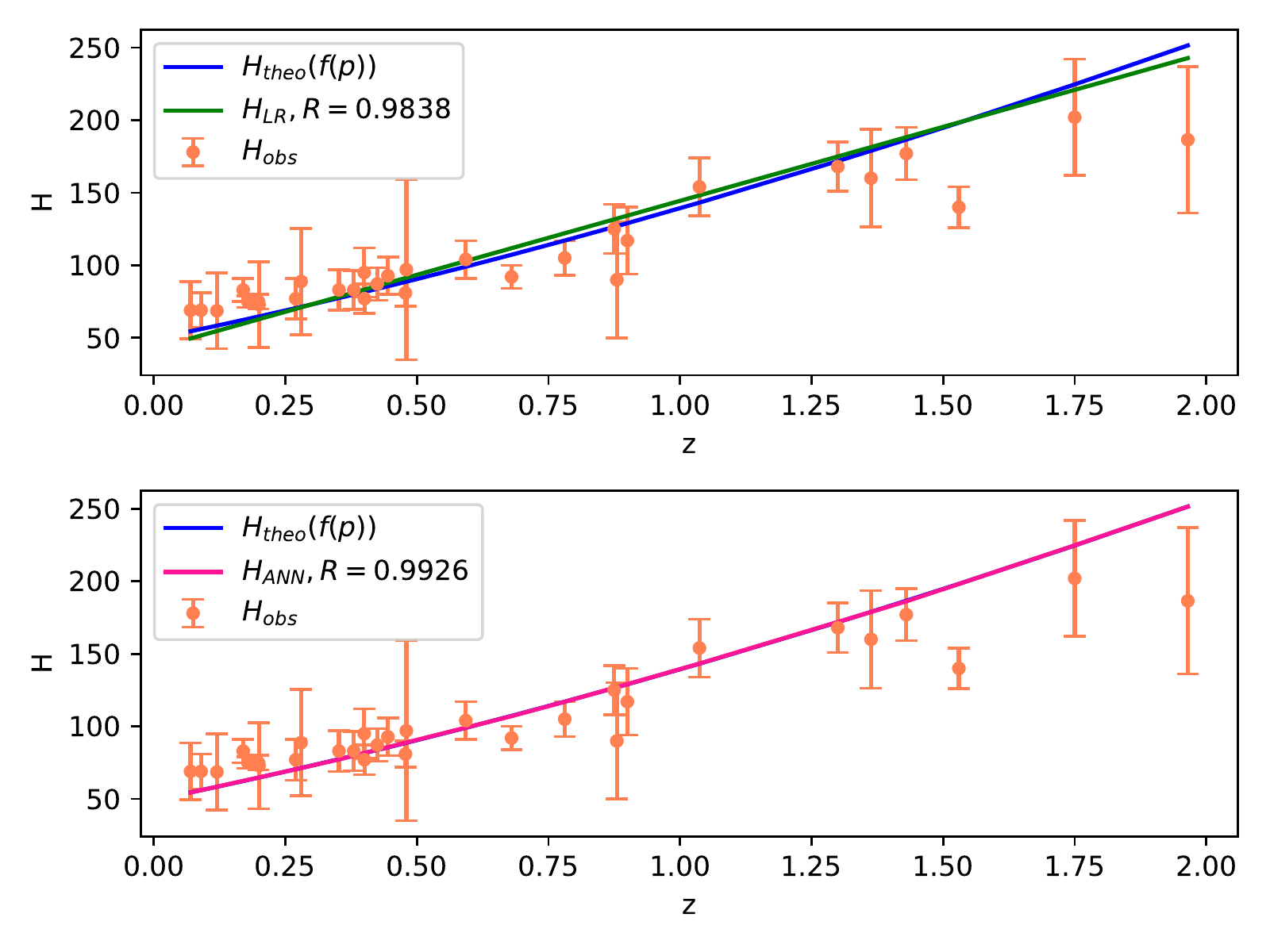}
\centering\caption{For $f(P)$ model: Plots of the estimated values of $H_{LR}(z)$ (top) and $H_{ANN}(z)$ (bottom) with the help of best fitted $H_{theo}$ and the observational $H_{obs}$ with error parameters.}
\label{Fig4a}
\end{figure}

\begin{figure}[hbt!]
\centering
\includegraphics{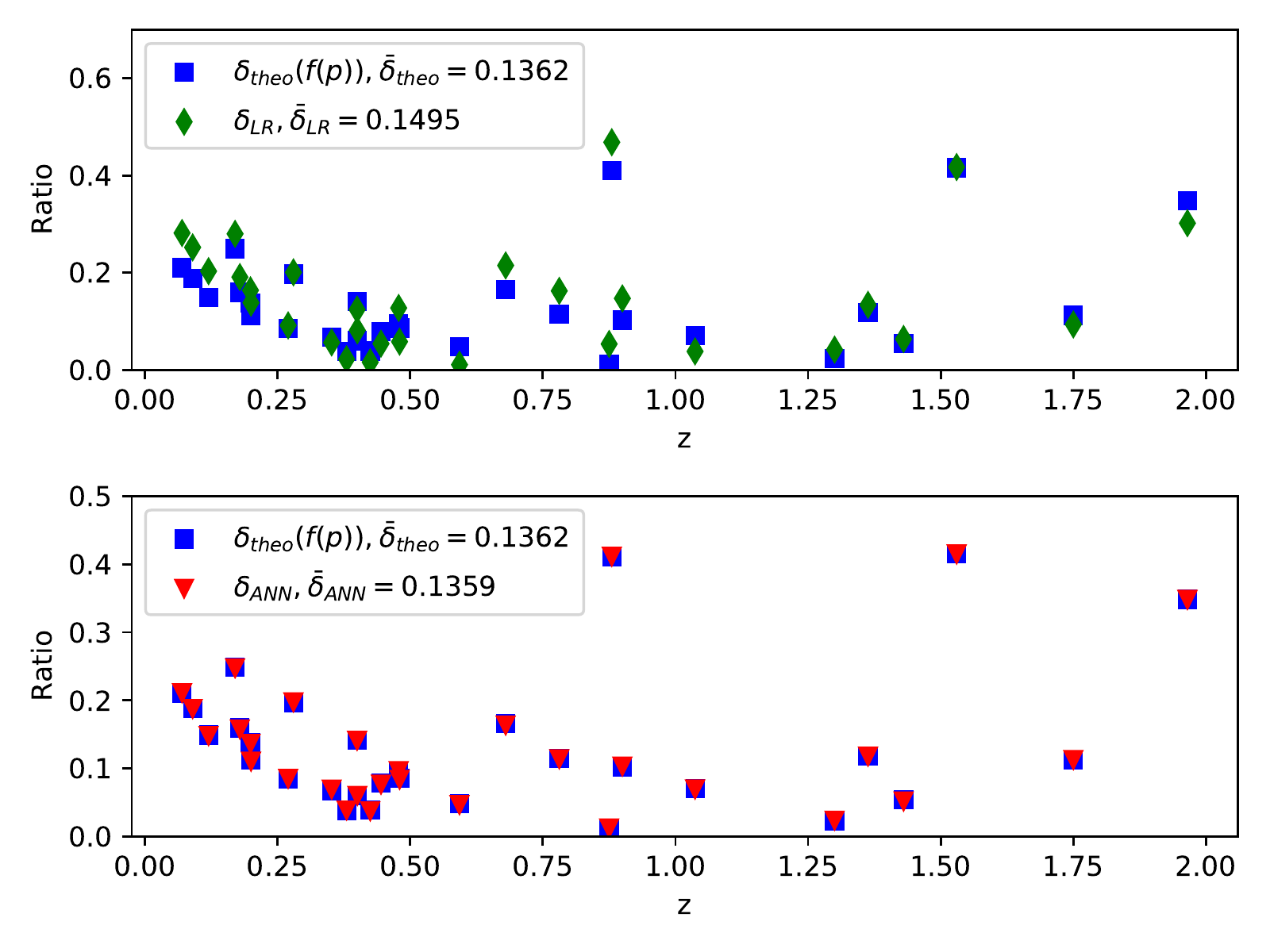}
\centering\caption{For $f(P)$ model: Comparing the deviation parameters ${\delta}_{theo}$ and ${\delta}_{LR}$ (top) and ${\delta}_{theo}$ and ${\delta}_{ANN}$ (bottom) with the help of best fitting values of the free parameters.}
\label{Fig4b}
\end{figure}

We have used multi-layer perceptron ANN to predict the values of $H$ which is denoted here as $H_{ANN}$. The estimated values $H_{ANN}$ are given in the column 5 of table \ref{MT} and \ref{MT1} for the respective theories of gravity. The lower panels of Fig. \ref{Fig3a} and \ref{Fig4a} show that the values of the correlation coefficient $R$ between the predicted $H_{ANN}$ and our theoretical models $H_{theo}$ are $0.9936$ (for cubic model) and $0.9926$ (for $f(P)$ model) respectively. So, it seems that ANN gives even better estimation of $H$ compare to the LR in both of the cases. We also calculate the ratios
$H_{ANN}/H_{obs}$ and the subsequent deviation parameter ${\delta}_{ANN}$ given by
\begin{equation}\label{Eqn1.5.185}
 {\delta}_{ANN} = |\frac{H_{ANN}}{H_{obs}} - 1|.   
\end{equation}
For, both of our models, in table  \ref{MT} and \ref{MT1}, $H_{ANN}/H_{obs}$ and ${\delta}_{ANN}$ are provided in columns 8 and 11 respectively. The mean deviations for ANN and our theoretical cubic gravity model are
\begin{equation}\label{Eqn1.5.184new}
(\bar{\delta}_{theo}, \bar{\delta}_{ANN}) = (0.1362, 0.1355),
\end{equation}
while, for $f(P)$ model, the same is given as
\begin{equation}\label{Eqn1.5.184anew}
(\bar{\delta}_{theo}, \bar{\delta}_{ANN}) = (0.1362, 0.1359).
\end{equation}
In the lower panels of Fig \ref{Fig3b} and \ref{Fig4b}, 
we plot the deviations ${\delta}_{theo}$ and ${\delta}_{ANN}$ along $z$ for the respective models. 
It is interesting to see that for both the cubic and $f(P)$ model, the mean deviations of ANN with theory are even lesser compared to the deviations of LR with theory. Thus, it can be seen that the ANN and the LR models perform significantly well and the ANN model is the better among the two. 


\begin{table}[h!]
\begin{tabular}{| c@{\hskip .15in}  c@{\hskip .15in}  c@{\hskip .15in}  c@{\hskip .15in}  c@{\hskip .15in}  c@{\hskip .15in}  c@{\hskip .15in}  c@{\hskip .15in}  c@{\hskip .15in}  c@{\hskip .15in}  c@{\hskip .15in} |}%
\hline 
\bfseries z & \bfseries $H_{obs}$ & \bfseries $H_{theo} (cubic)$ & \bfseries $H_{LR}$  & \bfseries $H_{ANN}$ & \bfseries $\frac{H_{theo}}{H_{obs}}$ & \bfseries $\frac{H_{LR}}{H_{obs}}$ & \bfseries $\frac{H_{ANN}}{H_{obs}}$ & \bfseries ${\delta}_{theo}$ & \bfseries ${\delta}_{LR}$ 
     & \bfseries ${\delta}_{ANN}$  \\ \\
\hline
\begin{filecontents*}{output.csv}
z,H_Obs,H_Theory,H_MLR,H_ANN,H_Th/H_obs,H_MLR/H_Obs,H_ANN/H_Obs,Dth,Dml,Dan
0.07,69,56.21,51.26,55.65,0.8147,0.743,0.807,0.1853,0.257,0.1934
0.09,69,57.8,53.38,57.4,0.8376,0.7736,0.832,0.1624,0.2264,0.1682
0.12,68.6,60.2,56.56,60.01,0.8776,0.8244,0.875,0.1224,0.1756,0.1252
0.17,83,64.28,61.85,64.38,0.7744,0.7452,0.776,0.2256,0.2548,0.2244
0.18,75,65.02,62.8,65.16,0.8669,0.8373,0.869,0.1331,0.1627,0.1312
0.2,75,66.68,64.92,66.91,0.8891,0.8656,0.892,0.1109,0.1344,0.1079
0.2,72.9,66.76,65.02,66.99,0.9158,0.8919,0.919,0.0842,0.1081,0.081
0.27,77,72.69,72.43,73.1,0.944,0.9407,0.949,0.056,0.0593,0.0507
0.28,88.8,73.55,73.49,73.97,0.8283,0.8276,0.833,0.1717,0.1724,0.167
0.35,83,79.85,81.11,80.25,0.962,0.9772,0.967,0.038,0.0228,0.0331
0.38,83,82.36,84.09,82.71,0.9922,1.0132,0.997,0.0078,0.0132,0.0035
0.4,95,84.14,86.19,84.44,0.8856,0.9073,0.889,0.1144,0.0927,0.1112
0.4,77,84.17,86.23,84.47,1.0931,1.1199,1.097,0.0931,0.1199,0.0971
0.42,87.1,86.37,88.8,86.59,0.9916,1.0196,0.994,0.0084,0.0196,0.0058
0.44,92.8,88.22,90.95,88.36,0.9507,0.9801,0.952,0.0493,0.0199,0.0478
0.48,80.9,91.29,94.48,91.27,1.1285,1.1678,1.128,0.1285,0.1678,0.1282
0.48,97,91.45,94.66,91.42,0.9428,0.9758,0.942,0.0572,0.0242,0.0575
0.59,104,102.12,106.62,101.28,0.9819,1.0252,0.974,0.0181,0.0252,0.0262
0.68,92,110.6,115.82,109.77,1.2022,1.259,1.193,0.2022,0.259,0.1932
0.78,105,120.72,126.51,120.42,1.1498,1.2049,1.147,0.1498,0.2049,0.1469
0.88,125,130.41,136.46,130.34,1.0433,1.0917,1.043,0.0433,0.0917,0.0427
0.88,90,130.93,136.99,130.86,1.4548,1.5221,1.454,0.4548,0.5221,0.4541
0.9,117,133.02,139.11,132.97,1.137,1.189,1.137,0.137,0.189,0.1365
1.04,154,147.67,153.61,147.69,0.9589,0.9974,0.959,0.0411,0.0026,0.0409
1.3,168,177.17,181.44,176.56,1.0546,1.08,1.051,0.0546,0.08,0.051
1.36,160,184.49,188.11,184.25,1.1531,1.1757,1.152,0.1531,0.1757,0.1516
1.43,177,192.39,195.2,192.43,1.087,1.1028,1.087,0.087,0.1028,0.0872
1.53,140,204.39,205.78,204.64,1.4599,1.4699,1.462,0.4599,0.4699,0.4617
1.75,202,231.59,229.07,232.21,1.1465,1.134,1.15,0.1465,0.134,0.1495
1.96,186.5,259.24,251.82,259.15,1.39,1.3503,1.39,0.39,0.3503,0.3895
\end{filecontents*}
\csvreader{output.csv}{}
{\\ \csvcoli & \csvcolii & \csvcoliii &\csvcoliv & \csvcolv & \csvcolvi & \csvcolvii & \csvcolviii & \csvcolix & \csvcolx & \csvcolxi} \\
\hline
\end{tabular}
\centering \caption{Numerical values of $H(z)$ for cubic gravity and the deviations with observed values}\label{MT}
\end{table}

\begin{table}[h!]
\begin{tabular}{| c@{\hskip .15in}  c@{\hskip .15in}  c@{\hskip .15in}  c@{\hskip .15in}  c@{\hskip .15in}  c@{\hskip .15in}  c@{\hskip .15in}  c@{\hskip .15in}  c@{\hskip .15in}  c@{\hskip .15in}  c@{\hskip .15in} |}%
\hline 
\bfseries z & \bfseries $H_{obs}$ & \bfseries $H_{theo} (f(P))$ & \bfseries $H_{LR}$  & \bfseries $H_{ANN}$ & \bfseries $\frac{H_{theo}}{H_{obs}}$ & \bfseries $\frac{H_{LR}}{H_{obs}}$ & \bfseries $\frac{H_{ANN}}{H_{obs}}$ & \bfseries ${\delta}_{theo}$ & \bfseries ${\delta}_{LR}$ 
     & \bfseries ${\delta}_{ANN}$  \\ \\
\hline
\begin{filecontents*}{outputfp.csv}
z,H_Obs,H_Theory,H_MLR,H_ANN,H_Th/H_obs,H_MLR/H_Obs,H_ANN/H_Obs,Dth,Dml,Dan
0.07,69,54.51,49.57,54.43,0.7899,0.7184,0.789,0.2101,0.2816,0.2111
0.09,69,56.04,51.61,56.03,0.8122,0.748,0.812,0.1878,0.252,0.1879
0.12,68.6,58.37,54.67,58.44,0.8509,0.7969,0.852,0.1491,0.2031,0.1482
0.17,83,62.32,59.77,62.44,0.7509,0.7201,0.752,0.2491,0.2799,0.2477
0.18,75,63.04,60.68,63.16,0.8406,0.8091,0.842,0.1594,0.1909,0.1578
0.2,75,64.65,62.72,64.76,0.8621,0.8363,0.864,0.1379,0.1637,0.1365
0.2,72.9,64.74,62.83,64.84,0.888,0.8618,0.889,0.112,0.1382,0.1105
0.27,77,70.48,69.96,70.45,0.9154,0.9086,0.915,0.0846,0.0914,0.0851
0.28,88.8,71.32,70.98,71.25,0.8031,0.7993,0.802,0.1969,0.2007,0.1976
0.35,83,77.42,78.32,77.23,0.9328,0.9437,0.93,0.0672,0.0563,0.0696
0.38,83,79.85,81.2,79.79,0.9621,0.9783,0.961,0.0379,0.0217,0.0387
0.4,95,81.58,83.22,81.59,0.8587,0.876,0.859,0.1413,0.124,0.1411
0.4,77,81.61,83.26,81.63,1.0599,1.0813,1.06,0.0599,0.0813,0.0601
0.42,87.1,83.75,85.74,83.84,0.9615,0.9843,0.963,0.0385,0.0157,0.0374
0.44,92.8,85.54,87.8,85.68,0.9218,0.9461,0.923,0.0782,0.0539,0.0767
0.48,80.9,88.52,91.2,88.71,1.0942,1.1273,1.097,0.0942,0.1273,0.0966
0.48,97,88.67,91.37,88.87,0.9141,0.942,0.916,0.0859,0.058,0.0838
0.59,104,99.02,102.9,99.15,0.9521,0.9894,0.953,0.0479,0.0106,0.0467
0.68,92,107.24,111.77,107.06,1.1657,1.2149,1.164,0.1657,0.2149,0.1637
0.78,105,117.06,122.06,116.97,1.1148,1.1625,1.114,0.1148,0.1625,0.114
0.88,125,126.45,131.65,126.55,1.0116,1.0532,1.012,0.0116,0.0532,0.0124
0.88,90,126.95,132.16,127.06,1.4106,1.4684,1.412,0.4106,0.4684,0.4117
0.9,117,128.99,134.2,129.09,1.1024,1.147,1.103,0.1024,0.147,0.1034
1.04,154,143.19,148.17,143.28,0.9298,0.9621,0.93,0.0702,0.0379,0.0696
1.3,168,171.8,174.98,171.98,1.0226,1.0416,1.024,0.0226,0.0416,0.0237
1.36,160,178.91,181.41,178.86,1.1182,1.1338,1.118,0.1182,0.1338,0.1179
1.43,177,186.57,188.24,186.17,1.0541,1.0635,1.052,0.0541,0.0635,0.0518
1.53,140,198.21,198.43,198.13,1.4158,1.4174,1.415,0.4158,0.4174,0.4152
1.75,202,224.63,220.86,224.78,1.112,1.0934,1.113,0.112,0.0934,0.1128
1.96,186.5,251.49,242.79,251.47,1.3484,1.3018,1.348,0.3484,0.3018,0.3484
\end{filecontents*}
\csvreader{outputfp.csv}{}
{\\ \csvcoli & \csvcolii & \csvcoliii &\csvcoliv & \csvcolv & \csvcolvi & \csvcolvii & \csvcolviii & \csvcolix & \csvcolx & \csvcolxi} \\
\hline
\end{tabular}
\centering \caption{Numerical values of $H(z)$ for $f(P)$ gravity model and the deviations with observed values}\label{MT1}
\end{table}

\section{Conclusion}
In this work we have performed an observational data analysis on the Einsteinian cubic gravity and $f(P)$ gravity. Both cubic and $f(P)$ gravity theories are higher order curvature theories that involve cubic corrections of the Riemann tensor to the Einstein Hilbert action. Motivated from their success as consistent theories of cosmology, we undertook the project of constraining their parameter space in this work. From the cosmological equations of the theories the theoretical framework was built by representing the Hubble parameter $H$ as a function of the redshift parameter $z$. The 30 point $z-H(z)$ cosmic chronometer data was used for the analysis. Clubbing the data with the additional constraints of BAO and CMB peaks further refinement was achieved in the probe. The Markov chain Monte Carlo method was used in the fitting analysis. Using our own PYTHON code and the publicly available CosmoMC code we found the best fit values of the free parameters of the model and also put bounds on them. 1D and 2D likelihood contours were generated for the free parameters representing their 68\% and 95\% confidence intervals. It was seen that the posterior distributions followed by the parameters are Gaussian-like, but quite skewed from a perfect normal distribution.

We went on to complement the fitting analysis with further statistical analysis and machine learning methods. We used correlation coefficients to compare the theoretical framework obtained from the fitting analysis with the observational data. For both cubic and $f(P)$ gravity it was found that the correlation coefficient was quite close to $+1$, showing strong positive correlation between theory and data. Two machine learning models, namely the linear regression and the artificial neural network was employed for the estimation of $H(z)$. We compared these estimates with the theory and the observations and it was found that these estimated models performed significantly well. Both of our models and the respective machine learning modelling predicted the behavior of $H(z)$ closely related to the observations. Further, the comparison of two machine learning models indicates that the ANN performs slightly better than LR. LR is method dealing with linear dependencies, ANN can deal with non-linearities. So, the data will have some nonlinear dependencies, ANN should perform better than regression. Interestingly, ANN is giving better forecasts compared to LR here. This clearly proves that the relation of z and H(z) is highly non-linear which is quite expected. Therefore, the studied models can be used to fill the data gaps of observational data sets that exist due to technological challenges and instrumentation constraints. This proves the accuracy of our fitting analysis and the consistency of the theory with the observations. This work is a significant theoretical development of the Einsteinian cubic and $f(P)$ gravity theories as far as their cosmological implication is concerned. It will be really interesting to check our results in the light of the results coming from larger data sets like Planck, Pantheon, etc., where the primary data sets are not $z-H(z)$ points, but involves measurements of other cosmological parameters. This may be an interesting future project for cubic and $f(P)$ gravity as far as observational constraints on the theoretical parameters are concerned.

\section*{Acknowledgments}
P.R. acknowledges the Inter University Centre for Astronomy and Astrophysics (IUCAA), Pune, India for granting visiting associateship. KG acknowledges the High Performance Computing System (HPC) at NITTTR Kolkata for using it as the computational resource for this paper. We thank the referee for his/her invaluable comments which helped us to improve the quality of the manuscript.




\end{document}